# Vacancy-mediated nitrogen diffusion and aggregation via high-fluence electron beam irradiation in HPHT synthesized diamond crystal

Chikara Shinei[1,2,*], Yuta Masuyama[3], Jun Chen[1], Hiroshi Abe[3], Masashi Miyakawa[1], Takashi Taniguchi[1], Takeshi Ohshima[3,4] and Tokuyuki Teraji[1]

[1] *National Institute for Materials Science, Tsukuba, Ibaraki 305 – 0044, Japan*

[2] *University of Tsukuba, Tsukuba, Ibaraki 305 – 0877, Japan*

[3]*National Institutes for Quantum Science and Technology, Takasaki, Gunma, 370 – 1292, Japan*

[4]*Department of Materials Science, Tohoku University Aoba, Sendai, Miyagi 980-8579, JAPAN*

[*]E-mail: shinei.chikara.fb@u.tsukuba.ac.jp

The negatively charged nitrogen vacancy ($NV^-$) center in diamond is a promising point defect for highly sensitive quantum sensing. The formation of high-density $NV^-$ centers is essential for improving sensitivity. We performed room-temperature electron beam irradiation (EBI) and annealing on nitrogen-doped high-pressure high-temperature diamond crystals, aiming to convert all substitutional nitrogen into NV structures by increasing EBI fluence. While the $N_s^0$ to $NV^{0/-}$ conversion process dominated at low EBI fluences, in a high-fluence region, $NV^{0/-}$ center and $N_s^{0/+}$ concentrations decreased with increasing EBI fluence, indicating the formation of unknown nitrogen-related defects such as the H3 center, which is a nitrogen and vacancy aggregation defect. Although H3 centers were observed at high EBI fluence, their annealing temperature of 1375 ± 25 °C was lower than the typically reported temperatures over 1600 °C. We attribute this low-temperature formation to vacancy-mediated nitrogen diffusion and aggregation.

## 1. Introduction

Negatively charged nitrogen-vacancy ($NV^-$) centers are paramagnetic defects with $S = 1$ in diamond crystals and have a strong spin-dependent photoluminescence intensity, which is promising for high-sensitivity quantum magnetic sensors.[1] Two factors that determine magnetic sensor sensitivity are the number of sensors, i.e., the $NV^-$ center concentration $[NV^-]$, and the spin coherence time $T_2$, which is the memory time of the sensing field strength as a phase of the quantum superposition state.[1] Based on these two factors, the shot-noise sensitivity is proportional to the equation[1]

$$\delta B \propto \sqrt{\frac{1}{[\mathrm{NV}^{-}] \times T_2}}. \quad (1)$$

Note that spin-dependent readout contrast (*C*), average number of photons collected per $NV^-$ center per measurement ($n_{avg}$), photon collection efficiency ($\beta$), and excitation diamond crystal volume (*V*) are also components of the shot-noise sensitivity[2]. Specifically, the spin-dependent readout contrast *C* depends on the excitation laser and microwave pulse schemes[3], while the other sensing parameters depend on the specific optical setup configurations. In this study, we specifically focus on the shot-noise components comprising intrinsic material parameters, as shown in Eq. (1). Furthermore, while the shot-noise sensitivity represented in Eq. (1) defines the classical limit for uncorrelated spin ensembles, quantum metrological techniques such as spin-squeezing[4] and spin-entanglement[5] have attracted increasing interest as a potential pathway to surpass the fundamental shot-noise limit toward the Heisenberg limit.

Shot-noise sensitivity is the minimum detectable magnetic field excluding various sensing system noises caused by environmental temperature fluctuations, excitation laser power fluctuations, and geomagnetic noise. As shown in Eq. (1), increasing the $T_2$ and $NV^-$ center concentrations improve the shot-noise sensitivity. Sekiguchi *et al.* reported that practical magnetic sensitivity corresponds to shot-noise sensitivity[6], as shown in Eq. (1) under condition of reduced $NV^-$ center excitation laser intensity noise. This practical sensitivity of 9.4 pT・$Hz^{1/2}$ was the best reported sensitivity in the low near-DC frequency range (5 to 100 Hz). In the higher near-DC frequency range above 100 Hz, practical sensitivities of 2.93 pT・$Hz^{1/2}$ [7] and 0.46 pT・$Hz^{1/2}$ [8] have recently been achieved. Sekiguchi *et al.* found that one of the reasons for this improved sensitivity was the elongation of the spin coherence time caused by the synthesis of high-quality diamond crystals. They also speculated that the best-reported sensitivity was limited by charge conversion from the $NV^-$ center to the neutrally charged NV center ($NV^0$). This charge conversion reduces the concentration of the activated $NV^-$ center, which is a quantum sensor, thus deteriorating shot-noise sensitivity.

It has been experimentally and theoretically reported that NV centers become negatively charged because of the electron supply from neutral substitutional nitrogen ($N_s^0$) to the $NV^0$ centers in diamond crystals.[9–12] In our previous study, we reported that the negative charge ratio $[NV^-]/[NV_T]$ ($[NV_T] = [NV^-] + [NV^0]$) was limited by the electron supply rate.[9] In this previous report, the NV centers, including negative and neutral charges, were formed by electron beam irradiation (EBI), subsequent annealing, and electron supply from $N_s^0$ to the $NV^0$ center, as shown in the following two reactions:

$$N_s^0 + V^0 \rightarrow NV^0 \ (2)$$

$$N_s^0 + NV^0 \rightleftarrows N_s^+ + NV^- \ (3)$$

EBI and annealing lead to the formation of neutral monovacancies ($V^0$) and $NV^0$ centers, as shown in Eq. (2). In diamond crystals used in practical sensing, the reported typical EBI fluence[6),13),14)] is up to ~ 1.0 × $10^{18}$ e/cm$^2$, and it leads to an $NV^-$ center density of a few parts per million. Therefore, a higher-fluence EBI is required to form a higher-density $NV^-$ center. However, we previously reported that increasing EBI fluence decreased the frequency of electron supply from $N_s^0$ to $NV^0$ centers because of the high frequency of $NV^0$ center formation.[9)]Although a 100% negative charge ratio for NV centers achieved using extra negative charging methods such as phosphorus doping to create an n-type diamond crystal[15),16)] and Fermi level control using diamond electronic devices,[17),18)] we aimed to achieve full conversion from $N_s^0$ to the NV structure by further increasing in EBI fluence.

In this study, we investigate the dependence of the $NV^{0/-}$ center concentration on EBI fluence under high-fluence EBI conditions of over 1 × $10^{18}$ e/cm$^2$ and discuss the effect of high-fluence EBI on the charge state of NV centers. Further, we discuss the correlation between $T_2$ and EBI fluence.

The organization of this article was as follows. In Sec. 3.1, we discuss formed NV center concentration depending on EBI fluence. Subsequently, we discuss formation of other unknown nitrogen-related defects except for $N_s$ and NV center (called $N_{other}$ in RESULT AND DISCUSSION) under condition of high-EBI fluence. The effect of $N_{other}$ on NV center charge state and the origin of $N_{other}$ are discussed in Sec. 3.2 and Sec. 3.3, respectively. Finally, we discuss spin coherence time of $NV^-$ center depending on the EBI fluence and the $N_{other}$ formation. Our findings are summarized in Sec. 4.

## 2. Methodology

The diamond crystals examined in this study were grown using a high-pressure high-temperature (HPHT) synthesis method. The nitrogen impurities in the HPHT diamond crystals primarily originated from the carbon source and the metal solvent used in the synthesis. Since natural nitrogen consists predominantly of $^{14}N$ (99.6%), this isotope was dominant in the synthesized crystals, and the concentration of [$^{13}C$] in the synthesized crystals was natural abundance 1.1 %. After the HPHT crystals were cut parallel to the {111} crystal plane, the top and bottom surfaces were mirror polished.

The NV centers were formed using the following procedure: The HPHT cutting diamond plate was irradiated using a 2.0 MeV electron beam with a total fluence range of $5 \times 10^{17}$–$3.6 \times 10^{18}$ e/cm$^2$, with subsequent annealing at 1375 ± 25 °C for 2 h in vacuum. Hereafter, we refer to the post process for NV center formation as "the e-beam/anneal process." The {111} growth sector was isolated by laser cutting to avoid underestimating the defect concentration in the HPHT diamond crystal because of multiple sectors. The dimensions of the {111} growth sector HPHT crystal used in this study were 2.0 × 1.0 × 0.4 mm$^3$. [$N_s^0$] and [$NV^-$] concentrations in the {111} growth sector before and after the e-beam/annealing process were estimated using electron paramagnetic resonance (EPR) measurements. EPR spectra were measured at ~300 K using an X-band EPR spectrometer (FA-100, JEOL). The microwave excitation frequency, microwave power, magnetic field modulation, and modulation amplitude were 9.428 GHz, 0.02 μW, 100 kHz, and 0.02 mT, respectively. The applied magnetic field was set parallel to the [111] direction of the cut diamond crystal. The estimated error in [$N_s^0$] and [$NV^-$] for the {111} growth sector was ~ ±10%. The concentration of $N_s^0$ formed during the HPHT synthesis, [$N_s^0$]$_{initial}$, was controlled by tuning the amount of titanium in the metal solvent and estimated as 6.8 ppm. The concentration of the NV centers formed during the HPHT synthesis was used as the detection limit for the EPR measurements in the examined HPHT diamond crystal. The details of $N_s^0$ and $NV^-$ center concentration estimation were explained in Supplemental section-1.

Photoluminescence (PL) measurements with 532 nm laser excitation were conducted to evaluate the $NV^0$ center concentration. A confocal micro-PL mapping system (Nanofinder FLEX; Tokyo Instruments, Inc.) was used to measure the PL intensity from the $NV^0$ and $NV^-$ centers. To estimate [$NV^0$], we multiplied the $NV^-$ center concentration obtained by EPR with the PL intensity ratio of the $NV^0$ center to the $NV^-$ center considering the difference in the PL decay rate between them. The details of $NV^0$ center concentration estimation were explained in Supplemental section-1. Ultraviolet PL (UVPL) measurements were conducted using home-built optical equipment that included a He-Cd laser (excitation wavelength = 325 nm) to evaluate the formation of nitrogen-related defects besides the substitutional nitrogen and NV centers.

Spin coherence time $T_2$ in the {111} growth sector HPHT diamond crystal was measured with columnar excitation fluorescence microscope[19)] and used to excite the $NV^-$ centers throughout the thickness direction of the diamond crystal (~ 0.4 mm) under a static bias magnetic field of 2.5 mT applied parallel to the [111] direction. The 532 nm laser diameter and depth of the excitation region in the employed microscope were designed to be 20 and 0.5 mm.[19),20)] The Hahn echo microwave and 532 nm laser protocol[19),20)] were implemented for the $NV^-$ centers in the excitation region to evaluate $T_2$. The sequence of the Hahn echo protocol is as follows: (1) A pulsed 532 nm laser with duration time of 180 μs was used to initialize the spin states of $NV^-$ centers. (2) Two sets of π/2 and one set of a π pulsed microwave, as determined by the period of Rabi oscillation for $NV^-$ center spins, were irradiated during the free precision time τ. We set the duration time of π/2 to be ~ 50 ns. (3) The examined HPHT diamond crystal was irradiated with a pulsed green laser with 5 μs duration time to read the $NV^-$ centers coherence state phase, and we detected the PL intensity emitted from the $NV^-$ centers. The PL intensity typically decays with increasing free-precision time, and the decay time corresponds to the spin coherence time. To measure the PL intensity depending on τ, S(τ), the Hahn echo protocol was conducted for various τ. For estimating $T_2$, we fit the experimentally obtained S(τ) to the expression $\exp[-(\tau/ T_2)^p]$, where the free parameters in the fit were $T_2$ and the stretched exponential parameter $p$. Our earlier work[19),20)] presented a detailed estimation scheme for the spin coherence time.

## 3. Results and Discussion

### 3.1 Effect of EBI fluence on formed NV center density

Figure 1(a) shows the dependence of defect concentration on EBI fluence. The defects indicate $N_s^0$, $N_s^+$, $NV^-$ centers, and $NV^0$ centers. As shown in Fig. 1(a), in the area where the EBI fluence is below $1.0 \times 10^{18}$ e/cm$^2$ (Area I), [$N_s^0$] decreased and both [$NV^0$] and [$NV^-$] increased. The rate of increase in [$NV^-$] with EBI fluence up to $1.0 \times 10^{18}$ e/cm$^2$ was ~0.13 ppm / $10^{17}$ e/cm$^2$. This value is comparable to the previously reported value of ~ 0.1 ppm/$10^{17}$ e/cm$^2$[21),22)] obtained for nitrogen-doped diamond crystals, wherein the $NV^{0/-}$ center creation process shown in Eq. (2) and (3) are dominant. The increase in the total NV center concentration [$NV_T$] = [$NV^0$] + [$NV^-$] between EBI fluences of $5 \times 10^{17}$ and $1 \times 10^{18}$ e/cm$^2$ was 1.5 ppm, and the decrease in the $N_s^0$ concentration was 1.8 ppm. The increase in [$NV_T$] and decrease in [$N_s^0$] were similar in magnitude, and therefore, the nitrogen-related defects present in the nitrogen-doped diamond after the e-beam/annealing process are dominated by $N_s^0$ and NV centers, as shown in Fig. 1(b). As shown in Fig. 1(a), in the area where the EBI fluence is greater than $1.3 \times 10^{18}$ e/cm$^2$ (Area II), the $N_s^0$, $NV^-$ center, and $NV^0$ center concentrations decreased with increasing EBI fluence. The disappearance of both the $N_s^0$ and NV centers led to the formation of other unknown nitrogen-related defects, $N_{other}$, as indicated in the schematic image for Area II in Fig. 1(b).

The objective of this study—full conversion from $N_s^0$ to the NV center—was not achieved. The formation of these unknown nitrogen-related defects may introduce acceptor levels into the diamond bandgap, thereby affecting the charge states of Ns and NV centers. Furthermore, these defects may generate magnetic noise for $NV^-$ centers, leading to degradation of $T_2$. Reducing the density of these defects is likely important for achieving high-sensitivity NV sensing.

### 3.2 Effect of $N_{other}$ on the charge state of NV centers

We previously reported that the $NV^-$ center concentration increases with an increase in the electron supply rate from $N_s^0$ to the $NV^0$ center when the e-beam/annealing process is performed[9),23)] This electron supply rate corresponds to the equilibrium constant $K$ in Eq. (3) and is defined as

$$K = \frac{[\mathrm{NV}^-][\mathrm{N_s^+}]}{[\mathrm{NV}^0][\mathrm{N_s^0}]}. \quad (4)$$

The equilibrium constant $K$ also limits the formation of $NV^0$ centers in addition to the $NV^-$ center and charge state ratio of $NV^-$ center to $NV^0$ center.[9),23)] Figure 1(c) shows the electron supply rate from $N_s^0$ to the $NV^0$ center, which is the equilibrium constant $K$. The applied $[N_s^0]$, $[NV^-]$, and $[NV_T]$ concentrations corresponded to the value presented in Fig. 1(a) for each EBI fluence. The black dotted line indicates $[NV_T]/[N_s^0] = [NV^-]/[N_s^0]$, which means that the formed NV centers are 100% negatively charged. The red triangle symbols represent the data of $\{[NV_T]/[N_s^0], [NV^-]/[N_s^0]\}$ obtained in this study. The gray square symbols represent the reported values of $\{[NV_T]/[N_s^0], [NV^-]/[N_s^0]\}$, in which electron exchange between the NV center and $N_s$ is dominant.[9)] As indicated by the red triangle symbol, $[NV^-]/[N_s^0]$ increased with an increase in $[NV_T]/[N_s^0]$ and the deviation between $[NV^-]/[N_s^0]$ and the fully negative charged line became large. This implies that the electron donor $N_s^0$ decreases and the $NV^0$ center increases with an increase in $[NV_T]$.[9)] The rate of increase in $[NV^-]/[N_s^0]$ relative to $[NV_T]/[N_s^0]$ corresponds to the equilibrium constant $K$ in the equilibrium equation shown in Eq. (3)[9)]; $K$ indicates the rate of electron supply from $N_s^0$ to the $NV^0$ center. For the nitrogen-doped HPHT diamond examined in this study, $K$ in Eq. (3) is expected to be smaller than the equilibrium constant for the gray triangle symbol in a previous study because of the acceptor levels associated with the formation of unknown nitrogen-related defects. The following equation for obtaining the negative charge ratio $F$ of NV centers was introduced to estimate the equilibrium constant $K$. In a previous study, this negative charge ratio equation was obtained from the equation of the equilibrium constant *K* shown in Eq. (4).[9),23)]

$$F = \frac{[\mathrm{NV}^-]}{[\mathrm{NV}^\mathrm{T}]} = \frac{-K\dfrac{[\mathrm{N_s^0}]}{[\mathrm{NV_T}]} + \sqrt{\left(K\dfrac{[\mathrm{N_s^0}]}{[\mathrm{NV_T}]}\right)^2 + 4K\dfrac{[\mathrm{N_s^0}]}{[\mathrm{NV_T}]}}}{2}. \tag{5}$$

Using this negative charge ratio equation, the vertical value of Fig. 1(c), namely, $[\mathrm{NV}^-]/[\mathrm{N_s}^0]$, can be expressed as a model function for $[\mathrm{NV_T}]/[\mathrm{N_s}^0]$ as

$$\frac{[\mathrm{NV}^-]}{[\mathrm{N_s^0}]} = \frac{[\mathrm{NV_T}]}{[\mathrm{N_s^0}]}\frac{[\mathrm{NV}^-]}{[\mathrm{NV_T}]} = \frac{[\mathrm{NV_T}]}{[\mathrm{N_s^0}]}\frac{-K\dfrac{[\mathrm{N_s^0}]}{[\mathrm{NV_T}]} + \sqrt{\left(K\dfrac{[\mathrm{N_s^0}]}{[\mathrm{NV_T}]}\right)^2 + 4K\dfrac{[\mathrm{N_s^0}]}{[\mathrm{NV_T}]}}}{2}, \tag{6}$$

where the fitting parameter was the equilibrium constant *K*. The fitting data of $\{[\mathrm{NV_T}]/[\mathrm{N_s}^0]$ and $[\mathrm{NV}^-]/[\mathrm{N_s}^0]\}$ was collected using Eq. (6) for the red and gray triangle symbol data. The obtained fitting parameter *K* for the red and gray triangle symbol data were 0.71 ± 0.08 and 0.53 ± 0.03, respectively. The obtained *K* for the red-triangle data was comparable to that for the gray-triangle symbol data. This result is inconsistent with our expectation of a significant decrease in *K* with the formation of unknown nitrogen-related defects. This result also shows that the charge states of the NV center and substitutional nitrogen are determined only by the equilibrium equation shown in Eq. (3). Unknown nitrogen-related defects are formed even if the number of NV centers and substitutional nitrogen atoms decreases.

The concentration of $[\mathrm{N_s}^+]$ was estimated using the *K* obtained by fitting Eq. (5). Based on the definition of equilibrium constant *K* in Eq. (3),[9),23)] $[\mathrm{N_s}^+]$ can be expressed as

$$[\mathrm{N_s^+}] = K\frac{[\mathrm{NV^0}][\mathrm{N_s^0}]}{[\mathrm{NV}^-]}. \tag{7}$$

We estimated $[\mathrm{N_s}^+]$ using the $[\mathrm{N_s}^0]$, $[\mathrm{NV}^-]$, and $[\mathrm{NV}^0]$ values obtained for each EBI fluence, as shown in Fig. 1(a) and Eq. (7). The estimated $[\mathrm{N_s}^+]$ values are shown as pink triangles in Fig. 1 (a). $[\mathrm{N_s}^+]$ tended to be slightly increased up to EBI fluence of $1.0 \times 10^{18}$ e/cm$^2$ and to decrease over an EBI fluence of $1.0 \times 10^{18}$ e/cm$^2$. The decrease in the $\mathrm{N_s}^+$ concentration indicates the conversion of $\mathrm{N_s}^+$ to unknown nitrogen-related defects. In addition, the $\mathrm{N_s}^+$ concentration was comparable to the $\mathrm{NV}^-$ concentration at any EBI fluence; this result indicates that strong binding between $\mathrm{NV}^-$ center and $\mathrm{N_s}^+$ caused by electron supply, and the electron supply from unknown nitrogen-related defects to the $\mathrm{NV}^0$ center is unlikely to occur. We estimated the unknown nitrogen-related defect concentrations by subtracting $[\mathrm{N_s}^0]$, $[\mathrm{N_s}^+]$, $[\mathrm{NV}^-]$, and $[\mathrm{NV}^0]$ obtained after the e-beam/annealing process from the as-grown nitrogen concentration $[N_s^0]_{initial}$ using

$$[N_{other}] = [N_s^0]_{initial} - ([N_s^{0/+}]+[NV^{0/-}]). \tag{8}$$

Figure 1(d) shows the dependence of $[N_{other}]$ on EBI fluence, as estimated using Eq. (8). As shown in Area I, $[N_{other}]$ is ~ 0 in the range of the EBI fluence, which indicates that no unknown nitrogen-related defects formed. In Area II, $[N_{other}]$ increases with an increase in the EBI fluence. At an examined maximum EBI fluence of $3.6 \times 10^{18}$ e/cm$^2$, $[N_{other}]$ reached ~ 4.5 ppm. This result confirms that ~70% of the total nitrogen concentration is converted to an unknown nitrogen-related defect.

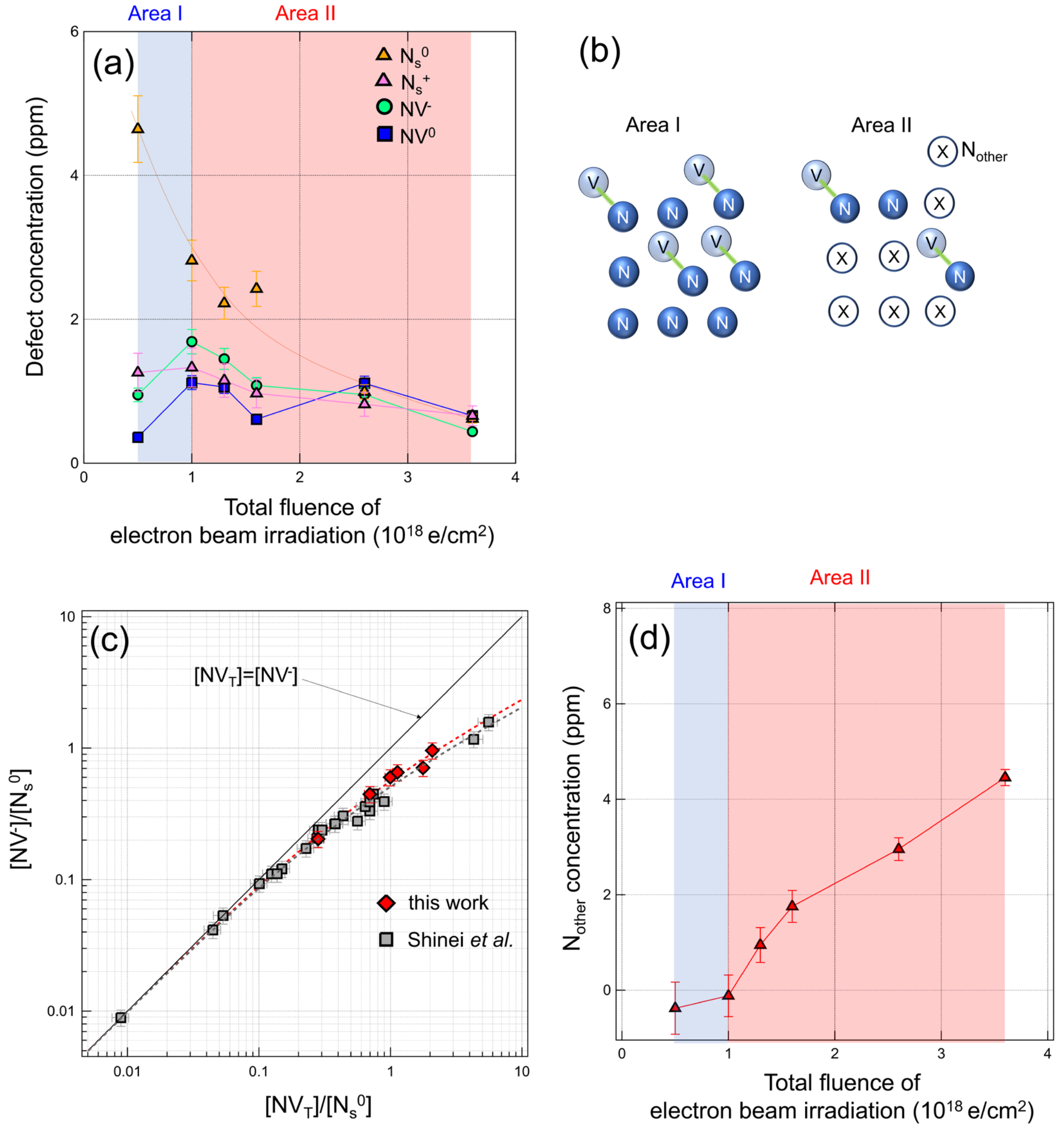

**Fig. 1.** (a) Dependence of EBI fluence on defect concentration; $[N_s^0]$, $[N_s^+]$, $[NV^0]$, and $[NV^-]$. The details of error bar estimation of these defects concentration for each EBI fluence

were explained in Supplemental section-1. (b) Dominant defects for two ranges of EBI fluence. Area I and Area II correspond to 5 × $10^{17}$–1 × $10^{18}$ e/cm$^2$ and 1 × $10^{18}$ e/cm$^2$–3.6×$10^{18}$ e/cm$^2$, respectively. (c) Reaction rate for the supply of an electron from $N_s^0$ to $NV^0$ center. The red and gray dotted line indicated fitting line using model function of Eq. (5) for this work data (red triangle symbol) and previous data (gray square symbol)[9] of {[$NV_T$]/[$N_s^0$], [$NV^-$]/[$N_s^0$]}. The error bars of [$NV_T$]/[$N_s^0$] and [$NV^-$]/[$N_s^0$] for each EBI fluence were obtained using error propagation rule with error of [$N_s^0$], [$NV^-$] and [$NV^0$] as shown in (a). (d) Dependence of EBI fluence on unknown nitrogen-related defects $N_{other}$ concentration. The error bar of [$N_{other}$] for each EBI fluence were obtained using error propagation rule with error of [$N_s^0$], [$N_s^+$], [$NV^-$] and [$NV^0$] as shown in (a). Solid lines as shown in (a) and (d) are eye guides.

### 3.3 Origin of $N_{other}$

The typical dominated charge state of the NV center and $N_s$ in the diamond crystal with the nitrogen concentration ranging from 1‐20 ppm was −/0[9)] and 0/+,[24)], respectively. The Fermi level in the nitrogen doped diamond located at 3.78 eV above the valence band level ($E_v$) [25] and charge transition level −/0 of the NV center and 0/+ of $N_s$ were $E_v$+2.7 eV and $E_v$+3.6 eV, respectively. Therefore, the increase in the $N_{other}$ concentration shown in Fig. 1(c) and the decrease in the $N_s^{0/+}$ and $NV^{0/-}$ concentrations shown in Fig. 1(a) suggest the conversion of defects with a single nitrogen atom to defects composed of multiple nitrogen atoms. Furthermore, identifying the defect structure of $N_{other}$, which is expected to be composed of multiple nitrogen atoms, is essential to suppress the decrease in the $NV^{0/-}$ center concentration. Multiple nitrogen complex defects in diamond crystals such as the H3 center (N-V-$N^0$), H4 center (3N-V-V-N), and N3 center (3N-V) are observed as photoluminescence (PL) signals at 503.5, 496.2, and 415.2 nm, respectively[26] UV excitation PL measurements were performed at an excitation wavelength of 325 nm for the nitrogen-doped diamond crystals examined in this study to observe the PL signals of multiple nitrogen complex defects with low-wavelength ZPL. Figure 2(a) shows the UV-excited PL spectra for an EBI fluence of 7 × $10^{17}$ e/cm$^2$, where the formation of $NV^{0/-}$ centers was dominant, and for an EBI fluence of 3.6 × $10^{18}$ e/cm$^2$, where the formation of $N_{other}$ was dominant. The ZPL of the $NV^0$ center at 575 nm and corresponding phonon side bands were observed in both EBI fluences, whereas the ZPL of the H3 center (N-V-$N^0$) at 503.5 nm and the corresponding phonon side bands were observed only in the EBI fluence of 3.6 × $10^{18}$ e/cm$^2$. Therefore, one of the origins of the unknown nitrogen-related defects is the H3 center. It should be noted that the relative PL spectral intensity in Fig. 2(a) under 325-nm excitation does not directly scale with the absolute defect concentration ratio shown in Fig. 1(a). First, despite comparable concentrations of $NV^-$ center and $NV^0$ center, the $NV^0$ center ZPL (575 nm) and its phonon sideband (PSB) overwhelmingly dominate the spectrum, while the $NV^-$ center ZPL (637 nm) is obscured. This is because high-energy 325-nm (3.81 eV) UV illumination efficiently promotes photo-ionization of $NV^-$ center into $NV^0$ center[11] ($NV^- + h\nu \rightarrow NV^0 + e^-$), optically enhancing the steady-state $NV^0$ center population during measurement. Second, regarding why the H3 PL emission (both ZPL at 503.5 nm and PSB) appears smaller than $NV^0$ center despite $N_{other}$ being the dominant defect, a comparison of Huang-Rhys factor (HRF) defined as $S = -\ln(I_{ZPL}/I_{total})$[27]—$S = 3.3$ ($I_{ZPL}/I_{total} \sim 3.7$ %)[27] for $NV^0$ center, $S = 4.0$ ($I_{ZPL}/I_{total} \sim 1.8$ %)[9] for $NV^-$ center and $S \sim 2.9$ ($I_{ZPL}/I_{total} \sim 5.5$ %) for H3 center[28]—indicates that the HRF difference alone cannot account for the smaller H3 ZPL peak height. Instead, two plausible factors may explain the reduced H3 emission: (i) Davies *et al.* noted that shorter excitation wavelengths increase the contribution of continuum background absorption relative to the H3 band[28]. While their study examined polarized luminescence, which prevents a direct one-to-one conclusion for our unpolarized setup, this background absorption at 325 nm is a plausible factor that suppresses the excitation efficiency of H3 centers. (ii) Non-radiative energy transfer from H3 centers to co-existing nitrogen-aggregate defects (such as A-centers) can induce luminescence quenching of H3 centers[26], which represents another plausible mechanism hindering the emission process.

The formation of the H3 center is attributed to the diffusion and aggregation of nitrogen atoms and vacancies with annealing at over 1600 °C.[29]In this study, the formation of H3 centers is attributed to annealing at ~1375 °C for 2 h. Under the condition that nitrogen atoms diffused by kicking out carbon atoms in diamond crystal, the diffusion coefficient and diffusion length of nitrogen atoms can be estimated using the equation[30]

$$D = 0.4 \exp\left(-\frac{6.8}{k_B T}\right) cm^2 s^{-1} \tag{9}$$

$$L = \sqrt{2Dt}\ cm \tag{10}$$

where the activation energy for nitrogen diffusion generated by kicking out carbon atoms was 6.8 eV, and $k_B$, $T$, and $t$ are the Boltzmann constant, absolute temperature (K), and annealing time (s), respectively. The resulting estimated diffusion length of the nitrogen atoms was ~0.03 nm. Subsequently, the average distance between the nitrogen atoms was estimated using

$$L = \sqrt[3]{\frac{3}{4\pi N}} \tag{11}$$

The nitrogen concentration in the diamond crystals examined in this study was ~6.8 ppm, and as per Eq. (11), the average distance between nitrogen atoms was estimated to be ~6 nm. The estimated nitrogen diffusion length was sufficiently smaller than the average distance between the nitrogen atoms. These estimation results confirm that the diffusion and aggregation process of nitrogen atoms attributed to kicking carbon atoms with the annealing process of ~1375 °C for 2 h can be ignored.

An EBI fluence over $1 \times 10^{18}$ e/cm$^2$ forms monovacancies with a concentration of over ~10 ppm.[23),31)] By using Eq. (11), the average distance between monovacancies with a concentration greater than 10 ppm was estimated to be smaller than 3 nm, which is shorter than the average distance between nitrogen atoms (6 nm) in the diamond crystal used in this study. Vacancies with closer proximity to the nitrogen atoms may cause significantly larger diffusion coefficient than that in the case of nitrogen atom diffusion mediated by carbon atoms. Jones *et al.* predicted that vacancy-mediated nitrogen diffusion would increase the diffusion coefficient.[30)]

The decrease in $N_s^0$ and $NV^{0/-}$ concentrations and the observation of H3 centers with an increase in EBI fluence suggest that the following H3 center creation process[31)] occurs for some $N_s^0$ and $NV^{0/-}$ centers by vacancy-mediated nitrogen diffusion.

$$NV^{0/-}+N_s^0 \Rightarrow NVN^0 \quad (12)$$

$$N_s^0+ N_s^0+V^0 \Rightarrow NVN^0 \quad (13)$$

Further, the formed H3 center can accept an electron from $N_s^0$ as shown by

$$N_s^0+NVN^0 \leftrightarrows N_s^+ +NVN^- \quad (14).$$

The additional electron supply from $N_s^0$ to $NVN^0$ (H3 center) is expected to decrease the rate of electron supply from $N_s^0$ to the $NV^0$ center according to the equilibrium constant $K$ defined in Eq. (4). However, as shown in Fig. 1(c), the $K$ value obtained for the diamond crystal with a high-density $N_{other}$ including the H3 center is comparable to $K$ obtained for the diamond crystal in which the $N_s$ and NV centers were the dominant defects. We previously reported that the electron supply from $N_s^0$ to $V^0$ occurred preferentially with respect to the electron supply from $N_s^0$ to the $NV^0$ center, and the formation of V caused a decrease in $K$.[23)]

In this previous study, we mentioned that the lower energy level of V than that of the NV center in the diamond bandgap was considered one of the reasons for the decrease in $K$. The energy levels from the valence band level ($E_V$) of the NVN, $N_s$, and NV center are 3.3,[32)] 3.6,[32)] and 2.7 eV[32)] as shown in Fig. 2(b). The electron supply from $N_s^0$ to $NV^0$ center was considered to be more dominant than the electron supply from $N_s^0$ to $NVN^0$ because the energy level of NVN was greater than that of NV center by 0.6 eV, as shown in Fig. 2(b).

Under the condition that NVN was the main defect of $N_{other}$, it was naturally assumed that a decrease in $K$ was unlikely to occur; therefore, the effect of the electron supply from $N_s^0$ to $NVN^0$ on $K$ was believed to be significantly small. The CL signal intensity of $NVN^0$ (H3 center) was significantly higher than that of $NVN^-$ (H2 center), as shown in Fig. SI-3 of the supplemental data. This large difference in the population of the NVN charge state may indicate a small frequency of electron supply from $N_s^0$ to $NVN^0$, as shown in Eq. (14) from the left to right. The inverse electron supply from $NVN^-$ to $Ns^+$ in Eq. (14) leading to $N_s^0$ formation is also unlikely to occur in the examined diamond crystal with $N_{other}$ including the H3 center because of the strong binding between $NV^-$ center and $N_s^+$ that is $[NV^-]$ limiting $[N_s^+]$ as discussed in Sec. 3.1. While the present study clarifies the vacancy-mediated nitrogen diffusion and aggregation process in HPHT synthesized diamond under high-fluence EBI, it is noteworthy that defect formation dynamics can strongly depend on the crystal synthesis method. For instance, chemical vapor deposition (CVD) grown diamonds exhibit a unique specificity in defect formation, such as the propensity to form vacancy-related complexes like NVH[24),33)] even in the as-grown state. This raises an important question of whether the nitrogen and vacancy aggregation process (e.g., H3 center formation) observed in our HPHT samples occurs similarly in CVD diamonds under comparable high-fluence EBI conditions. A comprehensive quantitative comparison of these defect formation mechanisms across different synthesis methods will be a dedicated subject of our future work.

The "vacancy mediated nitrogen diffusion and aggregation" caused a significant reduction of the $N_s^0$ (P1 center) concentration and high density $NVN^0$ (H3 center) formation in the examined HPHT diamond crystal. Umemoto *et al.* reported that X-ray excited radio-luminescence spectra including ZPL and the phonon-sideband of the H3 center can be used as diamond scintillators for α- and β-ray measurements[34)] In this report, a significantly low transmittance of less than 10% when using a commercial Ib-diamond crystal was considered a deterioration factor for the number of detected photons induced by α- and β-rays. The reported low transmittance was caused by light absorption originating from high-density nitrogen impurities with concentrations of less than 200 ppm. In addition, a P1 center with a higher concentration caused a larger absorption coefficient in the diamond crystal[35)] Therefore, in this study, the HPHT diamond crystal under the conditions of H3 center rich and P1 center poor is expected to contribute to a significant increase in the performance of diamond scintillators.

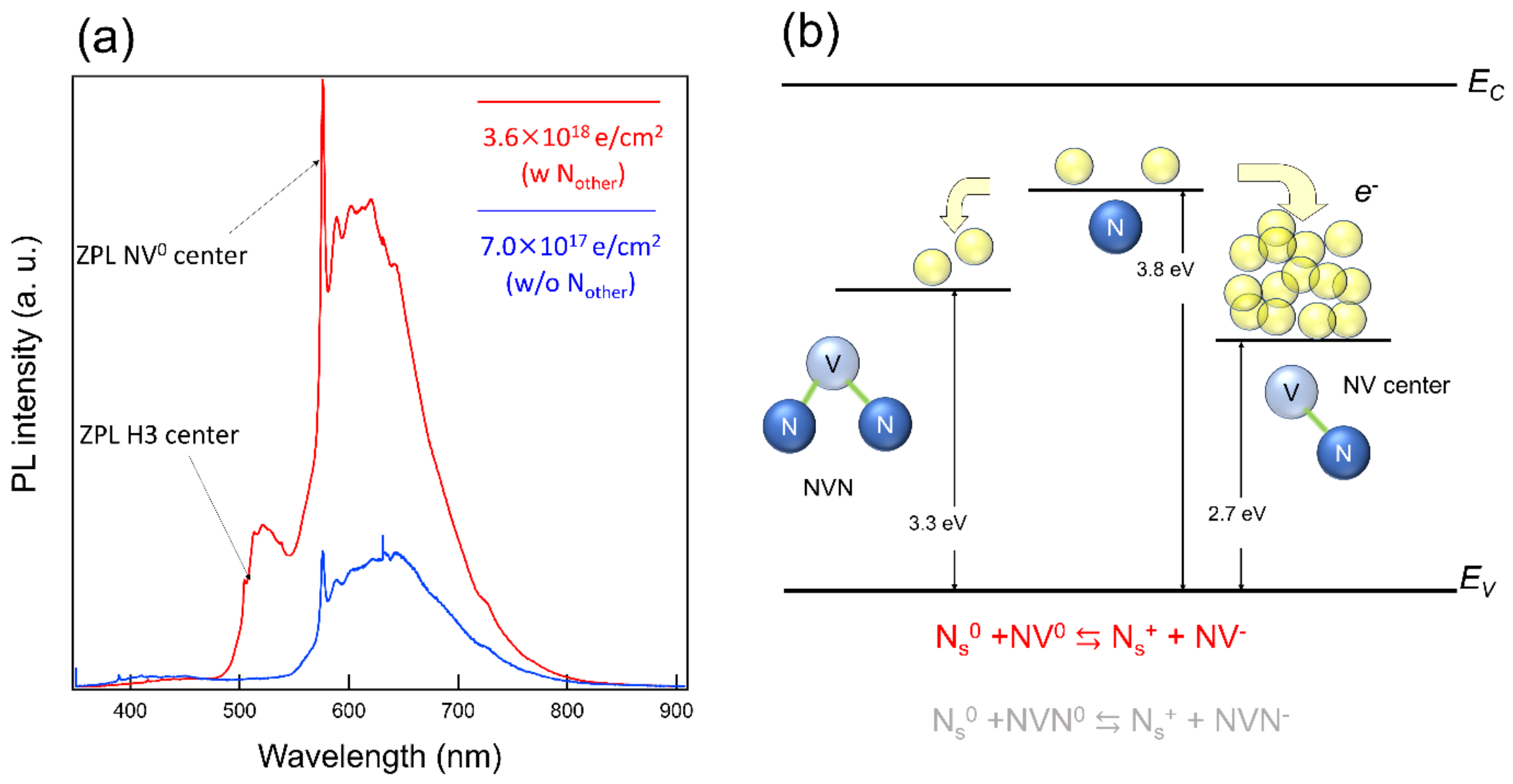

**Fig. 2.** (a) Ultraviolet photoluminescence spectrum for EBI influence of 7.0 × $10^{17}$ e/cm$^2$ (blue color) and 3.6 × $10^{18}$ e/cm$^2$ (red color). (b) Energy diagram in diamond band gap including $N_s$, NV center and NVN energy level. $E_c$ and $E_v$ indicate conduction and valence band levels, respectively.

### 3.4 Effect of EBI fluence on the spin coherence time $T_2$ of the NV$^-$ center

A Hahn-echo sequence, which utilizes a refocusing microwave pulse to reverse spin dephasing, was performed to investigate the NV$^-$ center spin decoherence caused by paramagnetic noise (the experimental details of Hahn-echo sequence are described in the METHODOLOGY section and the details of $T_2$ estimation are described in the Supplemental section-3). Through this sequence, DC noise—specifically the spin decoherence of the NV$^-$ center induced by the strain distribution in the diamond crystal—is cancelled. This allows for the selective extraction of AC decoherence driven by paramagnetic noise, such as that from P1 centers[20)]. Consequently, this sequence yields the spin coherence time, $T_2$, which is highly dependent on the paramagnetic noise density and serves as a critical indicator of NV sensing sensitivity as shown in Eq. (1). In this study, we investigated how the variation in defect concentrations induced by the electron beam irradiation (EBI) fluence affects this vital metric, $T_2$, thereby evaluating the robustness of the spin coherence.

Figure 3 shows the dependence of $T_2$ on the EBI fluence. At an EBI fluence of 1.0 × $10^{18}$ e/cm$^2$, the experimentally obtained $T_2^{exp.}$ (■) was a constant, i.e., it did not depend on EBI fluence. As shown in Fig. 1(a), the concentration of the NV$^{0/-}$ center, which is the decoherence source for the NV$^-$ center spin coherence, increased up to an EBI fluence of 1.0

× $10^{18}$ e/cm$^2$ and became comparable to the concentration of $N_s^0$, which is also the main decoherence source. In contrast, the decoherence contribution of the $NV^{0/-}$ centers to $T_2$ was comparable to that of $N_s^0$.[20]

Therefore, we discuss why $T_2$ remained constant despite the increase in the concentration of $NV^{0/-}$ centers. The concentrations of $NV^{0/-}$ centers formed by the e-beam and annealing processes and the concentration of $N_s^0$ not converted to $NV^{0/-}$ centers are denoted as [$NV^0$], [$NV^-$], and [$N_s^0$], respectively. We showed that, when only the formation process in which $N_s^0$ is converted to $NV^{0/-}$ centers occurs, $T_2$ can be estimated using the respective defect concentrations as [20]

$$\frac{1}{T_2^{\mathrm{sim}}} = D_{\mathrm{N_s^0}} \times \left[\mathrm{N_s^0}\right] + D_{\mathrm{NV^-}} \times [\mathrm{NV^-}] + D_{\mathrm{NV^0}} \times \left[\mathrm{NV^0}\right], \tag{15}$$

where $D_X$ represents the strength of the dipolar–dipolar interaction between the $NV^-$ center and paramagnetic defect X (X = $N_s^0$, $NV^-$ center, and $NV^0$ center). The applied $D_{Ns0}$, $D_{NV-}$, $D_{NV0}$ were 1/160 μs$^{-1}$ ppm$^{-1}$,[1),20),36)] 1/89 μs$^{-1}$ ppm$^{-1}$,[20] 1/160 μs$^{-1}$ ppm$^{-1}$,[20]. Using Eq. (15) and the EPR-estimated [$NV^0$], [$NV^-$], and [$N_s^0$] at each EBI fluence, we estimated the $T_2^{\mathrm{est.}}$ (●) shown in Fig. 4. Up to an EBI fluence of 1.0 × $10^{18}$ e/cm$^2$, $T_2^{\mathrm{est.}}$ (●) did not depend on the EBI fluence in the same manner as $T_2^{\mathrm{exp.}}$ (■). Furthermore, $T_2^{\mathrm{est.}}$ (●) was in agreement with $T_2^{\mathrm{exp.}}$ (■) within the fitting and estimated errors. This good agreement indicates that the EBI fluence-independent $T_2$ was reproduced by a model in which the $N_s^0$ and $NV^{0/-}$ centers were the dominant decoherence sources (Eq. (15)). Alsid *et al.* performed NV center formation using the e-beam and anneal process in a nitrogen-doped diamond with an as-grown formed nitrogen concentration of [$N_s^0$]$_{\mathrm{initial}}$ = 118 ppb, which was more than one order of magnitude lower than that in this study, and an observed EBI fluence-independent $T_2$ of ~ 420 μs at an EBI fluence below 1.0 ×$10^{18}$ e/cm$^2$.[31] In addition, we previously performed NV center formation using the e-beam and anneal process for CVD nitrogen-doped diamond with an as-grown formed nitrogen concentration: [$N_s^0$]$_{\mathrm{initial}}$ = 5 ppm and an observed EBI fluence-independent $T_2$ = 26 μs at an EBI fluence below 5.0 × $10^{17}$ e/cm$^2$.[20] From the abovementioned multiple previous studies and the present study, the $T_2$ independent of the EBI fluence was expected to be inversely correlated with the as-grown formed nitrogen concentration in a diamond crystal. We found that under conditions in which $NV^{0/-}$ center formation was dominant, $T_2$ was approximately determined by [$N_s^0$]$_{\mathrm{initial}}$ as

$$\frac{1}{T_2^{\mathrm{est.}}} \sim D_{\mathrm{N_s^0}} \times [\mathrm{N_s^0}]_{\mathrm{initial}}. \tag{16}$$

For details on the derivation of Eq. (16), please refer to the Appendix. Indeed, the EBI fluence-independent value of $T_2^{\mathrm{exp.}}$ (■) = 30 ± 8 μs obtained at an EBI fluence below 1.0 × $10^{18}$ e/cm$^2$ was comparable to the value of $T_2$ = 23 ± 2 μs obtained using Eq. (16) and $[N_s^0]_{initial}$ = 6.8 ppm of the diamond crystal examined in this study.

Subsequently, we investigated the effect of $N_{other}$ on the $T_2$ of the $NV^-$ centers. As shown in Fig. 3, $T_2^{exp.}$ (■) showed an increasing trend in an EBI fluence in the range of 1.3–2.6 × $10^{18}$ e/cm$^2$. In the same EBI fluence range, the $N_s^0$ and $NV^{0/-}$ center concentrations decreased with increasing EBI fluence as shown in Fig. 1(a). The main factor for the increase in $T_2$ in the range of the EBI fluence can be attributed to the decrease in the concentration of paramagnetic defects $N_s^0$ and $NV^{0/-}$ centers, which are decoherence sources. Above an EBI fluence of 2.6 × $10^{18}$ e/cm$^2$, $T_2^{exp.}$ (■) saturated to be ~ 50 μs, and $T_2^{est.}$ (●) increased monotonically. Furthermore, at an EBI fluence of 3.6 × $10^{18}$ e/cm$^2$, $T_2^{exp.}$ (■) was significantly smaller than $T_2^{est.}$ (●) that included the decoherence contribution of $N_s^0$ and $NV^{0/-}$ centers. The saturation $T_2^{exp.}$ (■) and the large discrepancy between $T_2^{exp.}$ (■) and $T_2^{est.}$ (●) at a high fluence of EBI suggested that paramagnetic defects of $N_{other}$ significantly increased above 2.6 × $10^{18}$ e/cm$^2$ and contributed to the decoherence of $T_2$ in addition to the decoherence contribution of the $N_s^0$ and $NV^{0/-}$ center.

To evaluate the practical sensing capability, we estimate the expected magneto-metric sensitivity in units of pT・$Hz^{1/2}$ based on the material figure-of-merit, $[NV^-]$ × $T_2$, which governs the shot-noise limit as shown in Eq. (1). In our sample irradiated at 1 × $10^{18}$ e/cm$^2$, the product $[NV^-]$ × $T_2$ reaches its maximum with $[NV^-]$ = 1.7 ppm (maximum $NV^-$ center concentration in this study) and $T_2$ = 30 μs obtained from Fig. 1(a) and Fig. 3. We benchmarked these material parameters against a well-established optical collection system reported by Wolf *et al.*[37),38)] (featuring an effective excitation volume of 8.5 × $10^{-4}$ mm$^3$ and a collection efficiency of 0.65, which achieved 9 pT・$Hz^{1/2}$ for $[NV^-]$ ~ 0.9 ppm and $T_2$ ~ 50 μs). Assuming for simplicity that differences in optically detected magnetic resonance (ODMR) contrast are negligible, applying our sample parameters to this measurement system yields an estimated Hahn-echo AC magnetic sensitivity of ~ 8.5 pT・$Hz^{1/2}$, which is comparable to the high sensitivity reported by Wolf *et al.* and represents a level among the highest sensitivities reported to date.

Furthermore, at this EBI fluence, $T_2$ is predominantly limited by Lorentzian magnetic noise from the paramagnetic electron-spin bath (such as P1 centers). Under such a spin-bath environment, applying dynamical decoupling (DD) with $N$ pulses is expected to extend the coherence time[39)] as $T_2(N) = T_{2,echo} \times N^{2/3}$. Although pulse control errors can cause experimental scaling to deviate from this theoretical limit[40)], combining uniform millimeter-scale microwave driving [8)](e.g., via dielectric resonators) with mains-frequency (50 Hz) noise synchronization[41] allows the system to closely approach ideal DD performance. Under the condition that the ideal XY16-16 sequence ($N$ = 256) is carried out for our diamond sample with EBI fluence of 1 × $10^{18}$ e/$cm^2$, the coherence time is expected to reach $T_2(256)$ ~ 1.2 ms, enhancing the expected AC magnetic field sensitivity to ~ 1.3 pT $Hz^{1/2}$ in the same optical setup[37),38].

It is also worth discussing the impact of $^{13}C$ nuclear spins in the HPHT diamond crystal, which contains natural isotopic abundance (1.1% $^{13}C$). As shown in the representative spin-echo decay curves in Supplemental Material (Fig. SI-3), coherent modulations due to Larmor precession of $^{13}C$ nuclear spins are observed. However, as demonstrated by Bauch *et al.*, the overall $T_2$ decay time in nitrogen-dominated regimes ([N] > 0.01 ppm) is independent of concentration, confirming that the EBI fluence dependence of $T_2$ in this work is governed by nitrogen-related defect spin dynamics. On the other hand, for DC magnetometry, the Ramsey dephasing time $T_2^*$ is severely limited to ~ 1μs by magnetic noise from 1.1% $^{13}C$ nuclear spins. Therefore, utilizing $^{12}C$-isotopically enriched HPHT diamond[2] is essential to significantly prolong $T_2^*$ and improve DC magnetic sensitivity, and investigating the EBI fluence dependence of $T_2^*$ in $^{12}C$ -enriched HPHT diamond remains an important study for future work.

Although the formation of $N_{other}$ as discussed in Sec. 3.2 and 3.3 and the depletion of $N_s^0$ donors leave a portion of the formed NV centers in the inactive neutral charge state ($NV^0$ center), combining EBI processing with optical pulse protocols offers significant potential to improve the effective charge-state conversion efficiency and sensing contrast. As noted in Section 1, the spin-dependent readout contrast ($C$) depends on the excitation laser and microwave pulse sequence schemes. In this context, applying a Spin-to-Charge Conversion (SCC) optical readout protocol—as demonstrated for NV ensembles by Jayakumar *et al.*[42] —presents a powerful sequence-based approach to significantly enhance $C$ and the signal-to-noise ratio (SNR) compared to conventional $NV^-$ center photoluminescence readout. The SCC protocol utilizes a relatively long 520-nm laser pulse (e.g., 1 ms) to initialize the charge state into $NV^-$ ($NV^0 + e^- \rightarrow NV^-$), followed by spin-selective photo-ionization pulses (e.g., via 594-nm laser excitation) that convert the spin projections of $NV^-$ center (spin sub level: $m_s = 0$ vs. $m_s = \pm1$) into distinct $NV^-/NV^0$ charge ratios prior to readout. Utilizing the $NV^0$ center population generated under high-fluence EBI via SCC sequence to boost $C$ complements the material optimization of $[NV^-] \times T_2$, providing a highly effective dual-approach toward achieving ultimate shot-noise-limited magnetic sensitivity.

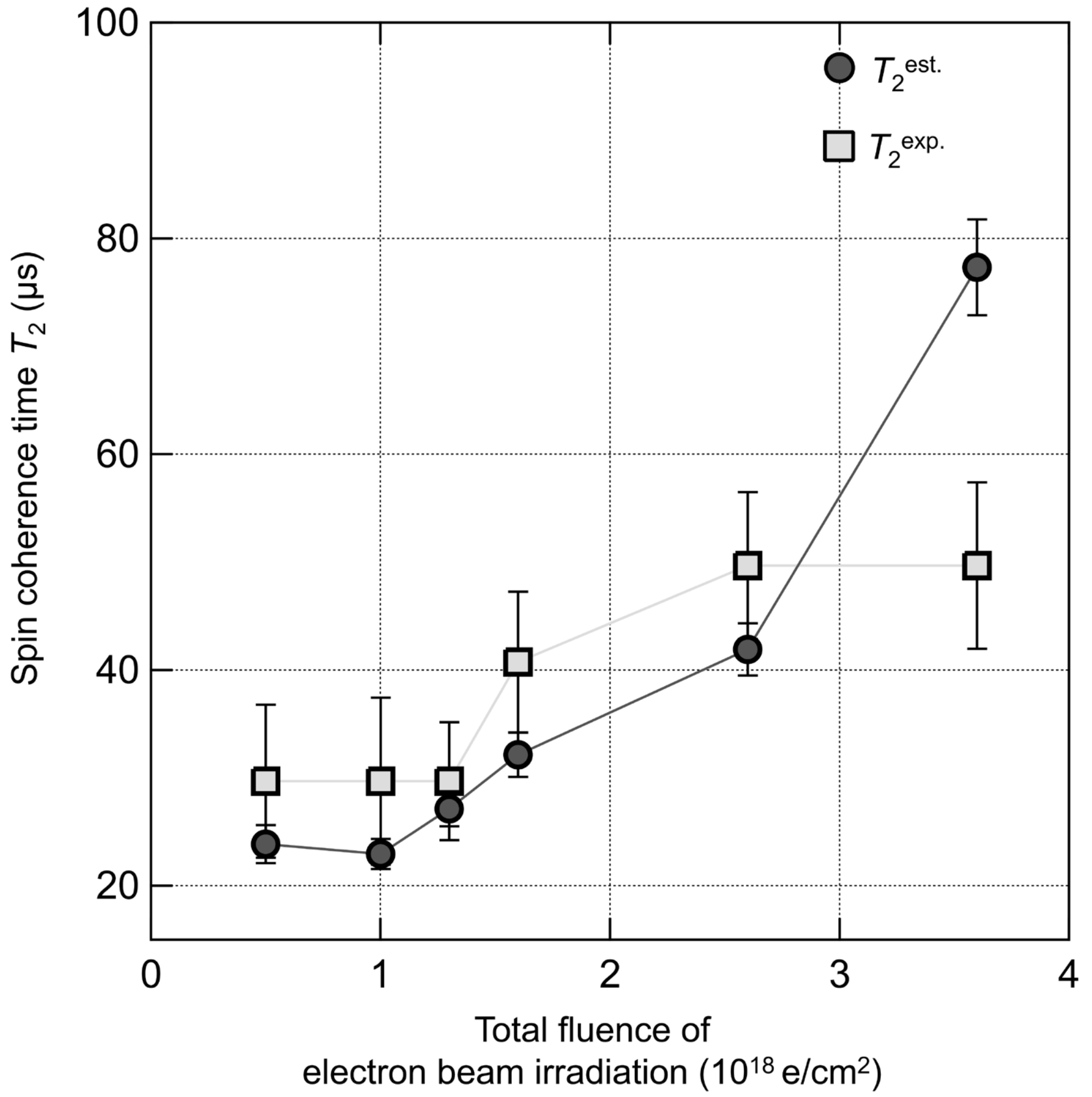

**Fig. 3.** Dependence of spin coherence time $T_2$ on EBI fluence. $T_2^{est.}$ and $T_2^{exp.}$, respectively show experimental and estimated values using Eq. (15). The vertical error bars of $T_2^{exp.}$ were obtained by the spin Hahn Echo signal S(τ) fitting. The vertical error bars of $T_2^{est.}$ were obtained using the estimation error of the defect concentration of $N_s^0$, $NV^0$ center, and $NV^-$ center.

## 4. Conclusion

In this study, we performed EBI at room temperature and annealed nitrogen-doped HPHT diamond single crystals to convert all substitutional nitrogen in the diamond crystal into NV structures. The concentration of as-grown nitrogen in the HPHT diamond single crystal used in this study was ~6.8 ppm. In the low EBI fluence region ($\leq 1.0 \times 10^{18}$ e/cm$^2$), the $N_s^0$ concentration decreased and $NV^{0/-}$ center concentration increased with increasing EBI fluence. The decrease in the neutral substitutional nitrogen ($N_s^0$) concentration was comparable to the increase in the total NV center concentration, which indicated that the conversion from $N_s^0$ to $NV^{0/-}$ centers was the dominant defect formation process in this region. In the high fluence EBI region (>1.0 × $10^{18}$ e/cm$^2$), both the $NV^{0/-}$ center and $N_s^0$ concentrations decreased with an increase in the EBI fluence. This result indicates the formation of unknown nitrogen-related defects other than $N_s$ and NV centers. The defect formation inhibited the conversion of all substitutional nitrogen into NV structures. Despite the formation of these unknown nitrogen-related defects, the charge state of the NV center was determined only by the electron supply rate from the $N_s^0$ to the $NV^0$ center, and some of these unknown nitrogen-related defects were nitrogen and vacancy aggregation defects, which are referred to as H3 centers ($NVN^0$). The H3 centers were observed in the high EBI fluence region, whereas the H3 forming annealing temperature of 1375 ± 25 °C was significantly lower than the typically reported temperatures over 1600 °C. We attribute the process of H3 center formation by high-fluence EBI and relatively lower annealing temperature to vacancy-mediated nitrogen diffusion and aggregation.

We also focused on the spin coherence time $T_2$ of the $NV^-$ center, which is a factor limiting magnetic sensitivity. When the correlation between EBI fluence and $T_2$ was examined, in the low EBI fluence range (≤ 1.0 × $10^{18}$ e/cm$^2$), $T_2$ did not depend on the EBI fluence and remained constant; this value was limited by the as-grown formed $N_s^0$ concentration. In the high fluence EBI range (> 1.0 × $10^{18}$ e/cm$^2$), $T_2$ showed an increasing trend as the fluence of EBI increased. Especially, over an EBI fluence of 2.6 × $10^{18}$ e/cm$^2$, the obtained $T_2$ saturated at ~ 50 μs. At the final EBI fluence of 3.6 × $10^{18}$ e/cm$^2$, the saturated $T_2$ was significantly smaller than the estimated $T_2$ including only the decoherence contribution of the $N_s^0$ and $NV^{0/-}$ centers. This result suggests that the concentration of paramagnetic defects other than $N_s^0$ and $NV^{0/-}$ centers significantly increased above 2.6 × $10^{18}$ e/cm$^2$ and contributed to the decoherence of $T_2$ in addition to the decoherence contribution of the $N_s^0$ and $NV^{0/-}$ centers.

## Acknowledgments

This work was supported by MEXT Q-LEAP (JPMXS0118068379 and JPMXS0118067395) and JST ASPIRE (JPMJAP24C1). T. Teraji acknowledges the support of JST CREST (JPMJCR1773), MIC R&D for the construction of a global quantum cryptography network (JPMI00316), and JSPS KAKENHI (Nos. 25K17636, 24K23011, 20H02187, 20H05661, and 19H02617). T. Teraji acknowledges the support of JST Moonshot R&D (JPMJMS2062).

## Appendix

**Derivation of Eq. (16)**

We studied the change in the spin coherence relaxation rate caused by the increase in the $NV^{0/-}$ centers concentration under the condition that the $NV^{0/-}$ centers formation was dominant, as shown in the EBI fluence range of Area I in Figs. 1(a) and (d). The reaction equations for the formation of $NV^{0/-}$ centers during the e-beam/annealing process are

$$N_s^0 + V^0 \rightarrow NV^0 \quad \text{(I)}$$

$$2\,N_s^0 + V^0 \rightarrow NV^- + N_s^+ \quad \text{(II)}$$

In Eqs. (I) and (II), three $N_s^0$ are used for the formation of one $NV^{0/-}$ center, and one $N_s^+$ is formed because of the electron supply from $N_s^0$ to the $NV^0$ center. From Eq. (I), the concentration of the formed $NV^0$ centers, $\Delta[NV^0]$, is equal to the concentration of $N_s^0$ used for the $NV^0$ center formation, $\Delta[N_s^0]_{NV0}$.

$$\Delta[\mathrm{N_s^0}]_{\mathrm{NV^0}} = \Delta\left[\mathrm{NV^0}\right] \quad \text{(III)}$$

From Eq. (II), the concentration of the formed $NV^-$ centers ($\Delta[NV^-]$) is equal to the half concentration of $N_s^0$ used for the $NV^-$ center formation ($\Delta[N_s^0]_{NV-}$) and concentration of $N_s^+$ ($\Delta[N_s^+]$).

$$\frac{1}{2} \times \Delta[\mathrm{N_s^0}]_{\mathrm{NV^-}} = \Delta[\mathrm{NV^-}] = \Delta[\mathrm{N_s^+}]. \quad \text{(IV)}$$

We divided the change in the spin coherence relaxation rate caused by the $NV^{0/-}$ center into $\Delta 1/T_2\{NV^0 \text{ formation}\}$ and $\Delta 1/T_2\{NV^- \text{ formation}\}$, and we expressed the spin coherence time $T_2$ after the formation of the $NV^{0/-}$ centers as

$$\frac{1}{T_2} = D_{\mathrm{N_s^0}} \times \left[\mathrm{N_s^0}\right]_{\mathrm{initial}} + \Delta\frac{1}{T_2}\left\{\mathrm{NV^0\ formation}\right\} + \Delta\frac{1}{T_2}\{\mathrm{NV^-\ formation}\}, \quad \text{(V)}$$

where $D_{Ns0}$, $[N_s^0]_{initial}$, and $C$ represent the DDI strengths of $N_s^0$, as-grown $N_s^0$ concentration, and $T_2$ of the single $NV^-$ center, respectively. $D_{Ns0}\times[N_s^0]_{initial}$ on the right-hand side of Eq. (V) indicates the $T_2$ of the as-grown $NV^-$ center.

Further, we estimated $1/T_2\{NV^0 \text{ formation}\}$. As shown in the following equation, the DDI strength of the $NV^0$ center was nearly equal to that of $N_s^0$ because the $N_s^0$ and $NV^0$ centers have the same spin number $S = 1/2$ and a similar electron gyromagnetic ratio of $g \sim 2.003$.[20]

$$\mathrm{D_{NV^0}} \sim \mathrm{D_{N_s^0}}. \quad \text{(VI)}$$

Using Eq. (IV) and (VI), we can describe $\Delta 1/T_2\{NV^0 \text{ formation}\}$ as

$$\Delta\frac{1}{T_2}\left\{\mathrm{NV^0\ formation}\right\} = (\mathrm{D_{NV^0}} \times \Delta\left[\mathrm{NV^0}\right] - \mathrm{D_{N_s^0}} \times \Delta\left[\mathrm{N_s^0}\right]_{\mathrm{NV^0}})$$

$$\sim(\mathrm{D_{N_s^0}} \times \Delta[\mathrm{N_s^0}]_{\mathrm{NV^0}} - \mathrm{D_{N_s^0}} \times \Delta\left[\mathrm{N_s^0}\right]_{\mathrm{NV^0}}) = 0. \quad \text{(VII)}$$

Equation (VII) shows that the increase in the spin relaxation rate caused by the $NV^0$ center formation is nearly equal to the decrease in the spin relaxation rate caused by the disappearance of $N_s^0$.

Further, we estimated $1/T_2\{NV^-$ formation$\}$. We previously reported that the DDI strength of $NV^-$ was approximately twice the DDI strength of the $N_s^0$ center because of the larger spin number of the $NV^-$ center ($S$=1) than that of $N_s^0$ ($S$ = 1/2).

$$\mathrm{D}_{\mathrm{NV}^-} \sim 2 \times \mathrm{D}_{\mathrm{N_s^0}}. \tag{VIII}$$

Using Eqs. (IV) and (VIII), we can describe $\Delta 1/T_2\{NV^-$ formation$\}$ as

$$\Delta\frac{1}{T_2}\{\mathrm{NV}^-\ \text{formation}\} = \mathrm{D}_{\mathrm{N_s^+}} \times \Delta[\mathrm{N_s^+}] + \left(\mathrm{D}_{\mathrm{NV}^-} \times \Delta[\mathrm{NV}^-] - \mathrm{D}_{\mathrm{N_s^0}} \times \Delta\left[\mathrm{N_s^0}\right]_{\mathrm{NV}^-}\right)$$
$$\sim \mathrm{D}_{\mathrm{N_s^+}} \times \Delta[\mathrm{N_s^+}] + \left(2 \times \mathrm{D}_{\mathrm{N_s^0}} \times \frac{1}{2} \times \Delta\left[\mathrm{N_s^0}\right]_{\mathrm{NV}^-} - \mathrm{D}_{\mathrm{N_s^0}} \times \Delta\left[\mathrm{N_s^0}\right]_{\mathrm{NV}^-}\right) = 0, \tag{IX}$$

where $D_{\mathrm{Ns+}} \times \Delta[N_s^+]$ on the right-hand side of Eq. (IX) is equal to 0 because $N_s^+$ is a diamagnetic defect.

From the results of $\Delta 1/T_2\{NV^0$ formation$\}$ and $\Delta 1/T_2\{NV^-$ formation$\}$ being 0, we indicated that the as-grown formed $N_s^0$ concentration $[N_s^0]_{initial}$ determined $T_2$ approximately as (corresponding to Eq. (16) in Sec. 3.4 of RESULT AND DISCUSSION)

$$\frac{1}{T_2} \sim D_{\mathrm{N_s^0}} \times \left[\mathrm{N_s^0}\right]_{\mathrm{initial}}. \tag{X}$$

**Supplemental Section-1: Defect Concentration Evaluation and Uncertainty Analysis**

The concentrations of [$N_s^0$] and [$NV^-$] were directly evaluated via double integration of 1st derivative X-band EPR spectra. Figure SI-1 (a) and (b) shows a representative 1st derivative EPR absorption spectrum and corresponding double-integration curve for a specific partial resonance line of $N_s^0$ (P1 center) and $NV^-$ centers, respectively (in the case of the examined HPHT diamond with an EBI fluence of $1.6 \times 10^{18}$ e/cm$^2$). Here, the evaluated line for the P1 center corresponds to the central $m_I$ = 0 hyperfine transition where the four N–C crystallographic orientation sites are degenerate, while that for the $NV^-$ center corresponds to the $m_s$ = -1 ⇔ $m_s$ = 0 transition for the [111]-oriented N–V axis site. Briefly, the double integration intensity of a specific EPR resonance line was multiplied by correction factors to account for the total defect population: a factor of 3 for the $N_s^0$ (P1 center) to account for the hyperfine splitting lines, and a factor of 8 for the $NV^-$ center to account for the fine structure splitting (2) and the four crystallographic NV axis orientations (4). Using the weight of the examined HPHT diamond crystal, the absolute defect concentrations of $N_s^0$ and $NV^-$ centers were estimated by calculating the ratio of these corrected intensities to that of a reference sample of $CuSO_4$・$5H_2O$. The error bars of these EPR-derived concentrations as shown in Fig. 1(a) are approximately ±10%, which mainly originates from analytical errors during baseline correction of the 1$^{st}$ derivative EPR absorption spectra. Figure S1 (c) presented representative 532 nm excited PL spectra (in case of examined HPHT diamond with EBI fluence of $1.6 \times 10^{18}$ e/cm$^2$), and corresponding Lorentzian fitting curve for ZPL of $NV^0$ center and $NV^-$ center were also shown. The concentration [$NV^0$] was determined from [$NV^-$] and the ZPL intensity ratio ($I_{ZPL}(NV^0)/I_{ZPL}(NV^-)$) corrected by the Huang-Rhys factors and radiative lifetimes (Eq.(1) in Appl. Phys. Lett. 119 (2021) 254001). The error bars for [$NV^0$] as shown in Fig. 1(a) were derived using the error propagation formula associated with the equation of relationship between [$NV^0$] and [$NV^-$] (Eq.(1) in Appl. Phys. Lett. 119 (2021) 254001), incorporating the ±10% EPR estimation error of [$NV^-$]. For full details on $N_s^0$, $NV^0$ and $NV^-$ center concentration estimation by EPR method and the PL method, see our earlier work of Appl. Phys. Lett. 119 (2021) 254001 and J. Appl. Phys. 132 (2022) 214402. For estimation of concentration of $N_s^+$ and its uncertainty, the charge-transfer equilibrium equation expressed as Eq. (7) ($[N_s^+] = K\frac{[NV^0][N_s^0]}{[NV^-]}$) was utilized. The error bars for [$N_s^+$] as shown in Fig. 1(a) were calculated using the error propagation formula associated with Eq. (7), incorporating the error of [$NV^0$], [$NV^-$], [$N_s^0$], and the fitting error of the equilibrium constant ($K = 0.71 \pm 0.08$).

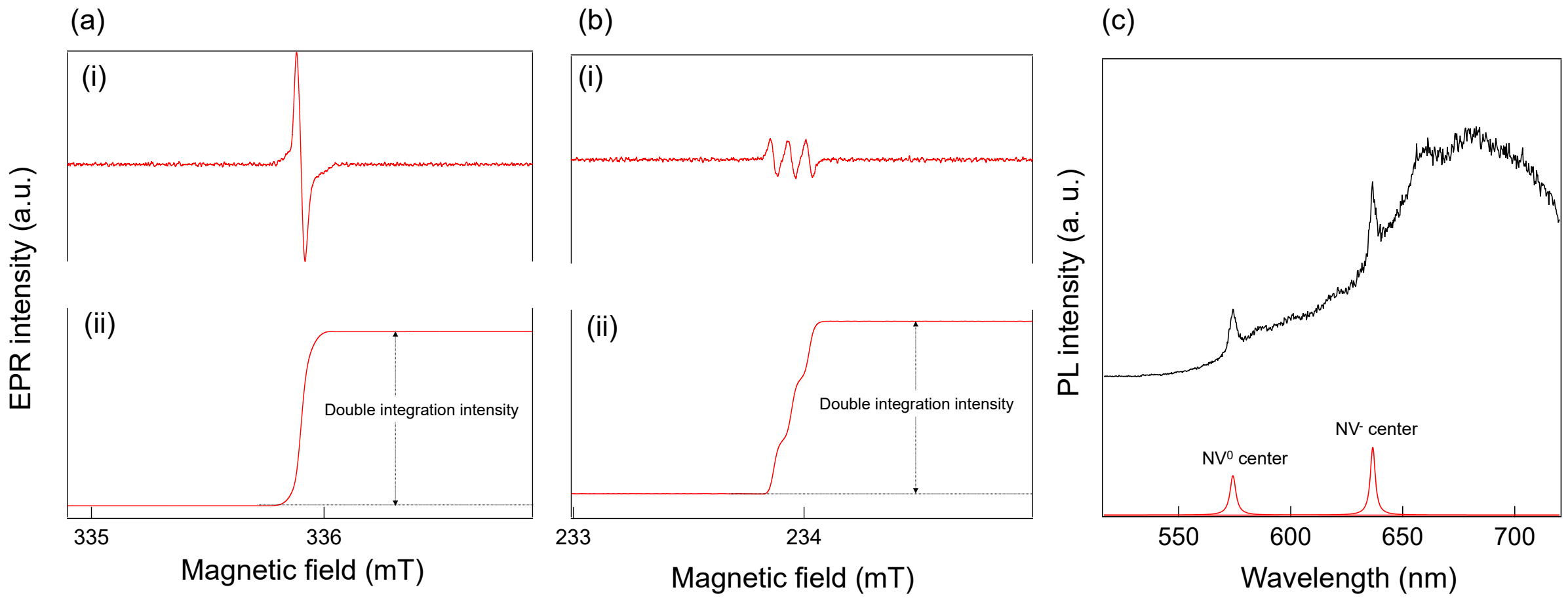

**Figure S1.** (a-i) 1st Derivative X-band EPR spectrum and (a-ii) corresponding double-integration curves for $N_s^0$ (P1 center) in examined HPHT diamond with EBI fluence of 1.6 × $10^{18}$ e/cm$^2$. (b-i) 1st Derivative X-band EPR spectrum and (b-ii) corresponding double-integration curves for $NV^-$ center in examined HPHT diamond with EBI fluence of 1.6 × $10^{18}$ e/cm$^2$. The three lines in the $NV^-$ center EPR spectrum correspond to $^{14}N$ hyperfine splitting. (c) 532 nm excited PL spectra including $NV^0$ center and $NV^-$ center PL emission (black solid curve) and corresponding Lorentzian fitting curve for ZPL of $NV^0$ and $NV^-$ center (red solid curve) in examined HPHT diamond with EBI fluence of 1.6 × $10^{18}$ e/cm$^2$.

## Supplemental Section-2: CL spectrum of nitrogen-doped HPHT diamond with high EBI fluence

We present the differences in the population of the NVN charge states. The CL measurements were performed by a HORIBA MP32 CL system attached to a Hitachi SU6600 field-emission scanning electron microscope (FE-SEM). The CL spectra in the UV-visible wavelength range (300~900 nm) were taken with a CCD detector at an accelerating voltage of 10 kV and a probe current of 7.2 nA; the CL spectra in the near-infrared wavelength range (900~1300 nm) were taken with an InGaAs detector at 15 kV and 76 nA. The CL measurements were performed at a cryogenic temperature of 80 K.

Figure S1 shows the CL spectrum of the examined nitrogen-doped HPHT diamond with an EBI fluence of 3.6 × $10^{18}$ e/cm$^2$. In the wavelength range of 300-900 nm, the band A structure (~ 440 nm), H3 center ZPL (504 nm) with the corresponding phonon sideband, and $NV^0$ center ZPL (575 nm) with the corresponding phonon sideband were observed. In the wavelength range of 900-1300 nm, the H2 center ZPL (987 nm) with weak intensity, H3 center second-order ZPL (1008 nm), and the $NV^0$ center second-order ZPL (1150 nm) were observed. The CL intensity of the H2 center was significantly lower than that of the H3 center, as shown in Figure S1. This large difference in the population for the NVN charge state may indicate a small frequency of electron supply from $N_s^0$ to $NVN^0$ in Eq. (14), as indicated in the original manuscript.

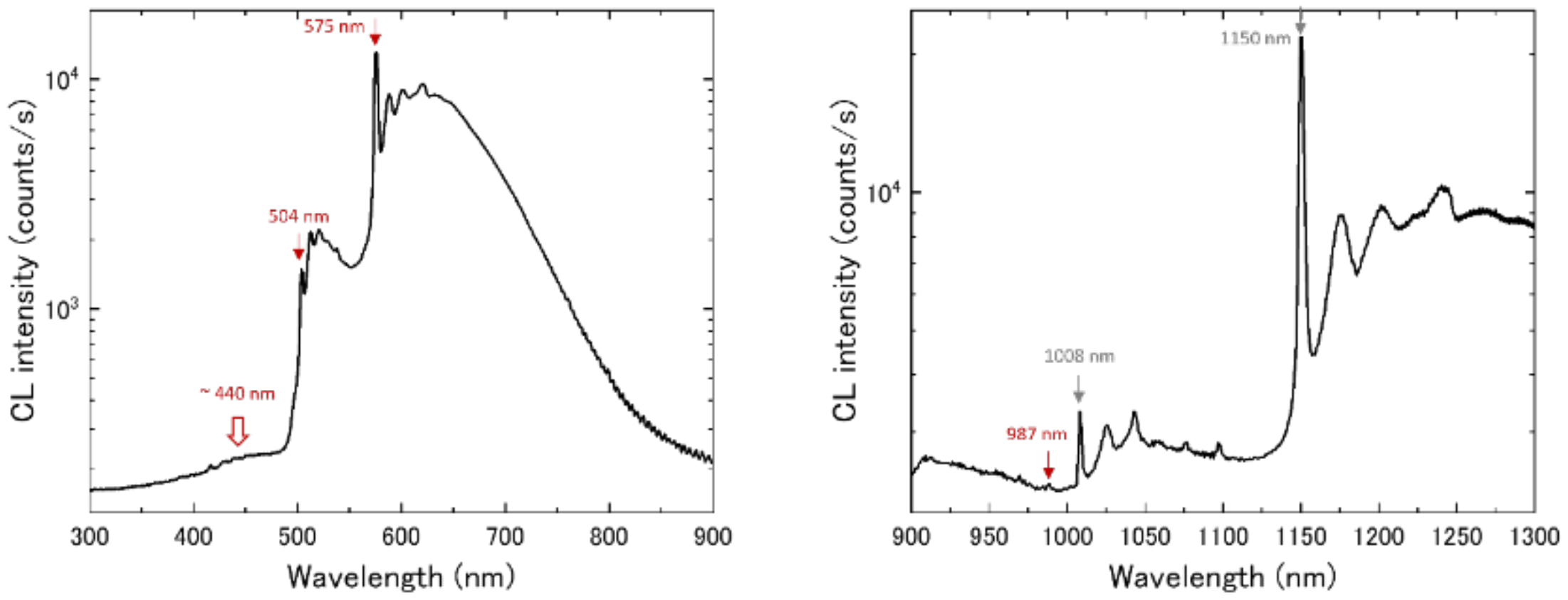

**Fig. S2.** CL intensity spectrum when using the nitrogen-doped HPHT diamond crystal in the original manuscript. The CL intensity of the H2 center is significantly weaker than that of H3 center. This may stem from two primary reasons: First, H2 centers typically form at high temperatures around 1600°C (J. Appl. Phys. 97 (2005) 083517.), whereas the annealing temperature in this study was below 1400°C, allowing H3 centers to dominate. Another consideration is that during CL observation with high-energy electron beam irradiation, H2 center may be ionized to form H3 center.

**Supplemental Section-3: Hahn-Echo Measurement and $T_2$ Envelope Fitting**

Hahn-echo measurements were conducted under a static bias magnetic field $\boldsymbol{B}$ of 2.5 mT applied parallel to the [111] diamond crystal direction. Figure S3 displays a representative Hahn-echo decay curve (in case of examined HPHT diamond with EBI fluence of 1.6 × $10^{18}$

e/cm$^2$). Under $\boldsymbol{B}$ of 2.5 mT, Larmor precession of $^{13}$C nuclear spins induces coherent collapse and revival modulations at intervals of ~ 73 μs (Phys. Rev B 85 (2012) 115303). The overall spin-coherence decay envelope connecting the $^{13}$C revival peak (blue circle) was fitted using a stretched exponential function $S(\tau) = \exp[-(\tau/T_2)^p]$ (red solid line), and the obtained $T_2$ and $p$ were 41 ± 6 and 1.2 ± 0.3, respectively.

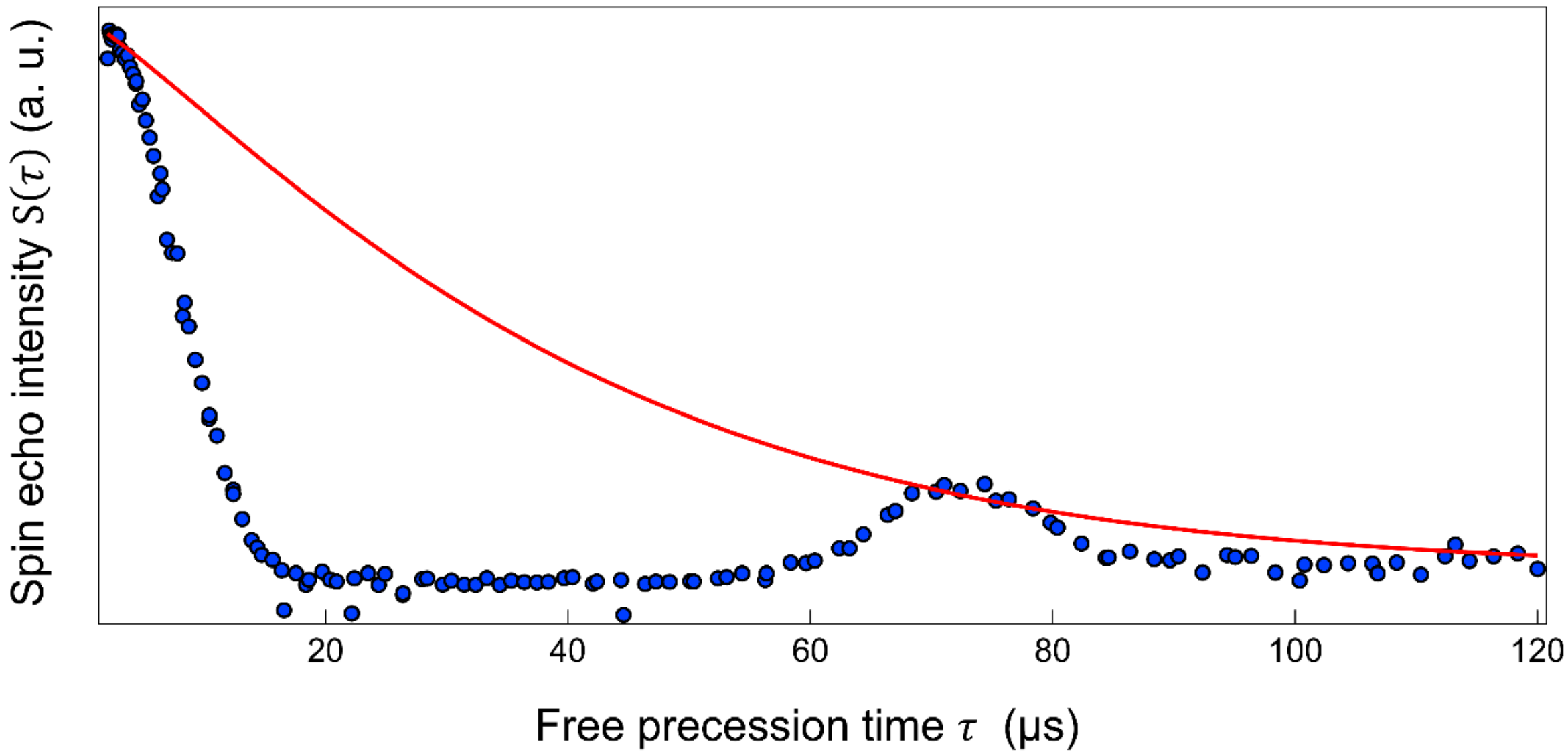

**Figure S3.** Representative Hahn-echo spin-decay curve in examined HPHT diamond with EBI fluence of 1.6 × 10$^{18}$ e/cm$^2$ under $\boldsymbol{B}$ = 2.5 mT // [111] with coherent modulations (a period of ~ 73 μs) originating from $^{13}$C nuclear spin Larmor precession (blue circle). The red solid line shows the fitted decay envelope ($S(\tau) = \exp[-(\tau/T_2)^p]$) used to extract $T_2$.